\documentclass[12pt]{article}

\usepackage{amsmath, amssymb}
\usepackage{graphicx,psfrag,epsf}
\usepackage{enumerate}
\usepackage{algpseudocode}
\usepackage{url} 
\usepackage[table]{xcolor}
\usepackage{booktabs}
\usepackage{calrsfs}
\DeclareMathAlphabet{\pazocal}{OMS}{zplm}{m}{n}

\usepackage{enumerate}
\usepackage{appendix}
\usepackage{tabularray}
\usepackage{caption}
\usepackage{subcaption}
\usepackage{xcolor}
\usepackage[linesnumbered, ruled, vlined, longend]{algorithm2e}
\RestyleAlgo{ruled}
\usepackage{bm}
\usepackage{adjustbox}
\usepackage{multirow}
\usepackage{comment}
\usepackage{enumitem}

\usepackage[hidelinks]{hyperref} 

\usepackage[
  backend=biber,
  style=apa,
  natbib=true,   
  url=true,
  doi=true,
  isbn=false,    
  eprint=false
]{biblatex}
\DeclareLanguageMapping{british}{british-apa} 

\newcommand{\blind}{0}

\newcommand{\Pois}{\mathsf{Pois}}
\newcommand{\E}{\mathbb{E}}
\newcommand{\Eq}{\E_{{q}_{\bm \phi}}}
\newcommand{\shp}{\mathsf{shp}}
\newcommand{\rte}{\mathsf{rte}}
\newcommand{\ELBO}{\mathsf{ELBO}}
\newcommand{\KL}{\mathsf{KL}}
\newcommand{\R}{\mathbb{R}}
\newcommand{\Rpos}{\R_{>0}}

\DeclareMathOperator{\argmin}{\mathrm{arg}\,\mathrm{min}\;}

\begin{document}

\def\spacingset#1{\renewcommand{\baselinestretch}%
{#1}\small\normalsize} \spacingset{1}

\if0\blind
{
  \title{\bf Seeded Poisson Factorization: leveraging domain knowledge to fit topic models}
  \author{Bernd Prostmaier\textsuperscript{1,2} \hspace{0.5em} Jan Vávra\textsuperscript{4} \hspace{0.5em} Bettina Grün\textsuperscript{3} \hspace{0.5em} Paul Hofmarcher\textsuperscript{2} 
  \vspace{2em}
  \date{\textsuperscript{1}BMW Group, Munich\\\vspace{0.5em}
  \textsuperscript{2}Department of Economics,\\
  Paris Lodron University Salzburg\\\vspace{0.5em}
  \textsuperscript{3}Institute for Statistics and Mathematics,\\
  WU Vienna University of Economics and Business\\\vspace{0.5em}
  \textsuperscript{4}Faculty of Mathematics and Physics,\\
 Charles University\\\vspace{0.5em}
  \today
  }
}
\maketitle
} \fi
\if1\blind
{
\title{Seeded Poisson Factorization: leveraging domain knowledge to fit topic model} 
\author{
  \vspace{2em}
  \date{
  \today
  }
}
\maketitle
} \fi

\begin{abstract}
Topic models are widely used for discovering latent thematic structures in large text corpora, yet traditional unsupervised methods often struggle to align with pre-defined conceptual domains. This paper introduces seeded Poisson Factorization (SPF), a novel approach that extends the Poisson Factorization (PF) framework by incorporating domain knowledge through seed words. SPF enables a structured topic discovery by modifying the prior distribution of topic-specific term intensities, assigning higher initial rates to pre-defined seed words. The model is estimated using variational inference with stochastic gradient optimization, ensuring scalability to large datasets.

We present in detail the results of applying SPF to an Amazon customer feedback dataset, leveraging pre-defined product categories as guiding structures. SPF achieves superior performance compared to alternative guided probabilistic topic models in terms of computational efficiency and classification performance. Robustness checks highlight SPF's ability to adaptively balance domain knowledge and data-driven topic discovery, even in case of imperfect seed word selection. 
Further applications of SPF to four additional benchmark datasets, where the corpus varies in size and the number of topics differs, demonstrate its general superior classification performance compared to the unseeded PF model.
\end{abstract}

\noindent%

\vfill

\newpage

\spacingset{1.45} 

\section{Introduction}\label{sec:introduction}

Inferring latent structures in text data is a fundamental challenge in natural language processing and its application in a wide range of fields of research such as political science, social science and economics. Due to the unstructured nature of text data, text analysis poses distinct challenges compared to the analysis of other types of data that are commonly used in empirical research \citep[see, e.g.,][]{text_selection}. Topic modeling provides a widely used framework for discovering hidden thematic structures within text corpora, offering insights into the distribution of topics across documents and the association between words and topics. Among the available topic modeling approaches, in particular Latent Dirichlet Allocation \citep[LDA;][]{10.5555/944919.944937} and its extensions \citep[see, e.g.,][]{NIPS2005_9e82757e, https://doi.org/10.1111/ajps.12103, eshima2023keyword}, which use the document-term matrix as input, have been widely studied and applied across various domains \citep[see, e.g.,][]{Bagozzi_Berliner_2018, BARBERÁ_CASAS_NAGLER_EGAN_BONNEAU_JOST_TUCKER_2019, words_numbers, 9581965, doi:10.1080/07350015.2020.1802285, ZIMMERMANN2024114310, LIU2025112905, CELIKTEN2025113219}. However, alternative topic modeling frameworks, such as Poisson Factorization (PF), provide distinct advantages by leveraging a Poisson likelihood rather than a multinomial distribution  and providing a more flexible prior parameter specification compared to LDA.

PF has been shown to provide a better fit to the data as well as improved scalability and computational efficiency \citep[see, e.g.,][]{GaP, 10.5555/2969033.2969181, gopalan2014scalable}. 
PF factorizes the document-term matrices into non-negative latent components, 
which correspond to topic intensities over words $\bm{\beta}$ (referred to as topical content or topic-term intensities), and document intensities over topics $\bm{\theta}$ (referred to as topical prevalence or document-topic intensities). The topical content refers to \emph{what} is being discussed, while the topical prevalence indicates \emph{how much} it is being discussed. PF naturally promotes sparsity and can handle large datasets efficiently due to its inherent properties and the use of variational inference techniques \citep[see, e.g.,][]{vafa-etal-2020-text,HofmarcherJAE,vavra_TPF_2024, vavra2024structuraltextbasedscalingmodel}.

Despite the success of topic models in uncovering latent themes in textual data, traditional methods are often limited by their purely unsupervised nature. In many applications, researchers and practitioners require models that align with pre-defined conceptual domains or that allow for targeted analysis based on domain knowledge \citep[see, e.g.,][for an application in political science]{eshima2023keyword}. Extensions of topic models to allow for guidance, e.g., via the inclusion of seed words, have thus been considered in a number of contributions, indicating their suitability to improve interpretability of topics as well as the use for automatic text classification. In this context, one can differentiate in particular between approaches extending non-probabilistic topic models, approaches extending the LDA-based probabilistic topic model and methods to improve the creation of seed words. 

The stream of literature extending non-probabilistic topic models includes among others
\citet{gallagher2017anchored} who pursue in their proposed anchored CorEx algorithm an information-theoretic framework which enforces a single-membership of words within topics and takes seed words into account as anchor words using as input a binarized version of the document-term matrix. Furthermore, exploiting the recent advances in large language models, \citet{pham2024topicgpt}  propose TopicGPT to uncover latent topics in a text collection based on a concise label paired with a broad one-sentence description to characterize a topic by prompting these models. \citet{grootendorst2022bertopic} combines in BERTopic document embeddings generated with pre-trained transformer-based language models with clustering of these embeddings and generating topic representations with a class-based term-frequency-inverse document frequency procedure, where an extension also allows for an inclusion of seed words.

LDA-based extensions of the probabilistic topic model to guide the topic estimation are considered by a number of contributions including \citet{10.5555/2380816.2380844, li2016seed, li2018seed, harandizadeh2022keyword, lin2023enhancing, watanabe, eshima2023keyword, watanabe2023}. 
These approaches typically influence the topic-term distributions, thereby enhancing interpretability and enabling more controlled modeling.

To improve guiding the estimation of topic models, approaches to derive a suitable set of seed words have also been investigated. These methods may be employed to obtain the seed words used as input to the guided topic model approaches and thus improve the overall performance. In this context, \citet{meng2020discriminative} propose the CatE approach, which learns a discriminative embedding space and discovers category representative terms in an iterative manner based on category names. \citet{zhang2023effective} propose their iterative framework SeedTopicMine, which allows them to jointly learn from three types of context suitable sets of terms to be used as seed words. 

Probabilistic topic models are built either on LDA or on PF. Although extensions of the LDA model to include guidance and seed words are numerous, we are not aware of work that extended PF in this direction, despite other extensions of the PF framework considered in the literature, such as for example \citet{duan2021topicnet} where the inclusion of a topic hierarchy is considered.  The paper at hand thus contributes to this literature by introducing a topic model using seed words within the PF framework. 
\color{black}

\begin{figure}[t!]
    \centering
        \includegraphics[width=0.8\textwidth]{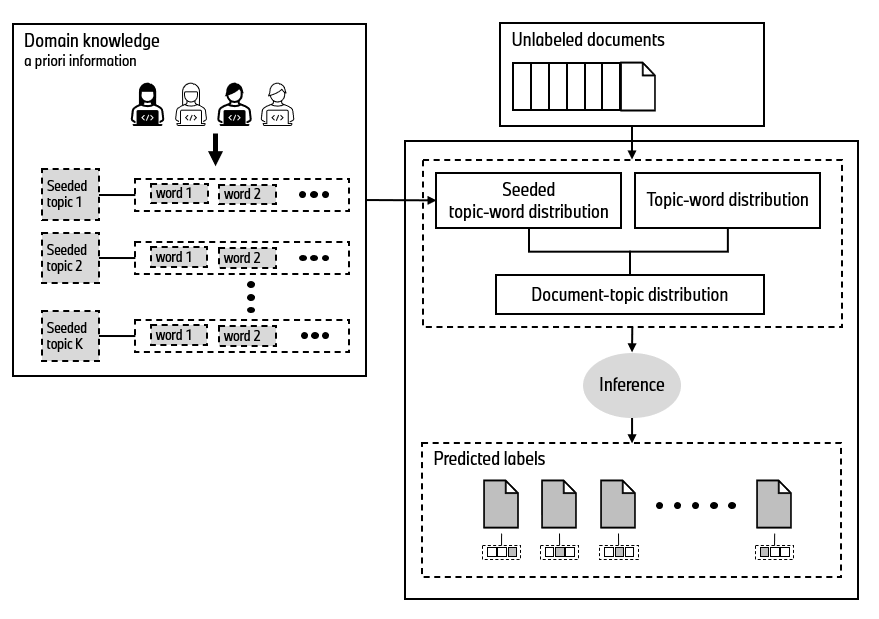}
    \caption{Architecture of the seeded Poisson Factorization (SPF) topic model. \label{fig:SPF_graphical_representation}}
\end{figure}

In particular, this paper contributes to the literature as follows. Firstly, we introduce \textit{seeded Poisson Factorization} (SPF), a novel topic model that integrates domain knowledge into the PF framework through the inclusion of seed words. As shown in Figure~\ref{fig:SPF_graphical_representation}, SPF extends standard PF by decomposing topic-term intensities into a neutral component and a seeded component informed by pre-defined seed words. This structured decomposition enables the model to incorporate prior knowledge while preserving the flexibility of PF to learn latent topics from data. In this way, SPF learns specific topics of interest and adaptively adjusts for potential seed word misspecifications by controlling the contribution of seeded components, ensuring robustness in applications where domain knowledge may be incomplete or imperfect. Secondly, in contrast to prior guided topic models, which predominantly rely on Markov Chain Monte Carlo (MCMC) inference \citep[see, e.g.,][]{eshima2023keyword, watanabe2023}, SPF employs variational inference (VI) for scalable parameter estimation. While MCMC methods are in principle applicable, we rely on VI methods, which formulate posterior inference as an optimization problem, significantly reducing computational costs compared to traditional sampling-based methods \citep{ranganath2013black, Blei_2017}. We empirically demonstrate that SPF not only achieves competitive predictive performance but also exhibits substantial computational efficiency, making it particularly suitable for large-scale text corpora.

Topic models are used for automatic text analysis in a wide range of fields of research, including for example 
text classification in data journalism \citep{10.1214/12-AOAS618}, 
open-ended survey responses in the social sciences \citep{https://doi.org/10.1111/ajps.12103} and analysis of speech data in  political science \citep{vavra2024structuraltextbasedscalingmodel}. In the following, we focus on yet another area of application: automatic analysis and classification of consumer feedback. Consumer feedback provides valuable insights into customer preferences, sentiment, and emerging trends \citep[see, e.g.,][]{7860894, AGUWA2017136, FILIERI2018956, 9581965, aghakhani2021, ZHANG2021107135, BISWAS2022113799, ZHOU2024114088}. Given the growing volume of online reviews and their impact on decision-making, accurately categorizing and summarizing this feedback remains a central challenge in computational social science and business analytics. Thus, we make use of a publicly available Amazon customer review dataset to illustrate and evaluate the performance of SPF. In particular, we assess SPF's performance to extract meaningful topics when seed words are supplied to characterize the underlying product categories of products discussed in the customer reviews. In addition, we investigate how well SPF infers the product category of the product discussed in a consumer review by employing a Naive Bayes classifier on the inferred topic intensities. We also compare the predictive performance as well as the computational efficiency of SPF to competing recently proposed guided topic models with readily available software implementations, i.e., KeyATM \citep{eshima2023keyword} and SeededLDA \citep{watanabe2023}. We conduct a series of robustness checks to examine the sensitivity of SPF to variations in seed word quality, model specification and corpus characteristics. In addition, we also fit SPF to four publicly available corpora with known categories to indicate the general applicability of our model for automatic text classification. Our results confirm that SPF provides excellent performance aunder various experimental conditions. 

By introducing SPF, we contribute to the methodological literature on topic modeling by extending PF with domain-informed priors, providing an alternative to existing LDA-based guided topic models. Our results demonstrate that SPF enhances both the guidance and computational efficiency of topic models, offering a scalable solution for researchers and practitioners seeking structured topic discovery in large text corpora.

The rest of the paper is structured as follows. In Section~\ref{sec:SPF}, we describe our generative model. Section~\ref{sec:inference} outlines the model inference. 
Section~\ref{sec:empirical} presents empirical results based on the application of SPF to Amazon customer feedback data as well as four benchmark datasets consisting of text corpora and their categorization.
Section~\ref{sec:conclusion} concludes.

\section{The seeded Poisson factorization model}\label{sec:SPF}

Based on the bag-of-words assumption \citep[see, e.g.,][]{eshima2023keyword} the data are summarized in a~Document-Term-Matrix (DTM), $\mathbb{Y}$. This matrix has the dimension number of documents $D$ times number of unique terms (words) in the data $V$, where each row corresponds to a~single document~$d = 1, \ldots, D$, and each column represents a~specific term $v = 1, \ldots, V$ from vocabulary $\mathcal{V}$. A~single entry $y_{dv}$ contains the frequency count of term $v$ in document $d$, such that $y_{dv}\ge 0$. PF topic models assume that the observed word frequencies are generated independently from a~Poisson distribution. The Poisson rates are decomposed into a~linear combination of document-topic intensities $\bm \theta$ and topic-term intensities $\bm \beta$ over latent topic dimension $K$ for every frequency count $y_{dv}$: 
\begin{equation*}
    y_{dv} \sim \Pois\left(\sum_{k=1}^K \theta_{dk}\beta_{kv}\right).
\end{equation*}
Document-topic intensities form a~tall matrix $\bm \theta = (\bm \theta_d)_{d=1}^{D} = (\theta_{dk})_{d,k=1}^{D,K}$, while topic-term intensities form a~wide matrix $\bm \beta = (\bm \beta_k)_{k=1}^{K} = (\beta_{kv})_{k,v=1}^{K,V}$. The number of topics $K$ needs to be a-priori specified. Both intensity matrices consist of positive elements. 

In its standard form, the PF models the frequency of words in documents without including any prior knowledge about the topic structure. However, in many applications, researchers possess domain knowledge that suggests certain terms that are highly indicative of specific topics. To leverage this information, we extend the standard PF by introducing structured priors based on the inclusion of seed words. Specifically, SPF differs from PF in that the topic-term intensities are decomposed into two components: a neutral component representing unsupervised topic discovery and a seeded component that emphasizes pre-identified important terms for specific topics. This structured decomposition allows the model to prioritize seed words during inference, steering the learned topics towards meaningful, interpretable structures aligned with user-specified domain knowledge.

In particular, we include the prior knowledge in SPF in the following way. We \emph{seed} the topics by inflating the prior mean of topic-term intensities for \emph{seeded} words. Let $\mathcal{V}_k \subset \mathcal{V}$ be the set of seed words for topic~$k=1,\ldots, K$ of size $V_k = | \mathcal{V}_k |$. In practice, we expect only a few seed words per topic, $V_k \ll V$ and denote by $\mathcal{S} = \bigcup_{k=1}^K \mathcal{S}_k$, $\mathcal{S}_k = \{(k,v), v\in \mathcal{V}_k\}$, the set of all \emph{seed words}. We allow for $\mathcal{V}_k = \emptyset$, $V_k = 0$, in which case the topic is not a-priori seeded. Note that in case $V_k = 0$ for all $k = 1, \ldots, K$, the SPF model reduces to a standard PF model.   

For the topic-term intensities we assume that they can be decomposed into a component present for all terms and a component specific to seed words, i.e., $\beta_{kv} = \beta_{kv}^\star + \widetilde{\beta}_{kv}$ where $\widetilde{\beta}_{kv} > 0$ for seeds and $\widetilde{\beta}_{kv} = 0$ otherwise. 
Both components are given a gamma prior:
\begin{equation} \label{eq:prior_betas}
    \beta_{kv}^\star \sim \Gamma(a,b)
    \quad \text{and} \quad
    \widetilde{\beta}_{kv} 
    \left\{
        \begin{aligned}
            &\sim \Gamma(c,d) && \;\text{for}\; (k,v) \in \mathcal{S}, \\
            &= 0 && \;\text{otherwise},
        \end{aligned}
    \right.
\end{equation}
with $a, b, c, d > 0$ and $c \gg a$. In case $b = d$, this implies that $\beta_{kv}$ also has a gamma prior $\Gamma(a + c, b)$ if $(k,v) \in \mathcal{S}$. All document-topic intensities $\theta_{dk}$ are given a gamma prior
\begin{equation} \label{eq:prior_theta}
    \theta_{dk} \sim \Gamma(e, f).
\end{equation}

\begin{figure}[ht!]
    \centering
    \includegraphics[width=0.8\textwidth, trim = 0 0 0 0, clip]{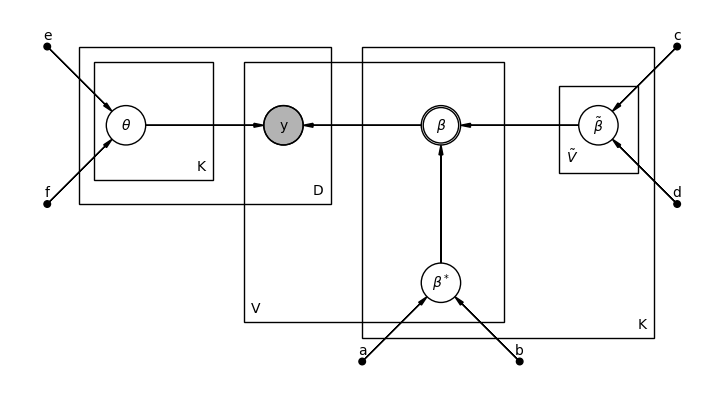}
    \caption{Directed graphical representation of the SPF model. Shaded nodes are observed, transparent nodes are latent variables, double circles indicate deterministic transformations of parent nodes and points are fixed parameters. \label{fig:DGM_SPF}}
\end{figure}

To obtain the empirical results in Section~\ref{sec:empirical}, we set $a = b = e = f = 0.3$, which according to \citet{gopalan2014scalable} results in sparse representations of the document-topic and topic-term intensities. To emphasize the relevance of the pre-defined seed words, we set $c = 1.0$ and $d = 0.3$. Finally, the generative process is captured in plate notation in Figure~\ref{fig:DGM_SPF}.

\section{Inference}\label{sec:inference}

\subsection{Variational inference}\label{sec:VI}

Given the DTM $\mathbb{Y}$, we infer the document-topic intensities and the topic-term intensities based on approximating the posterior distribution over the model's latent variables $p(\bm \theta, \bm \beta^\star, \widetilde{\bm \beta} \,|\, \mathbb{Y})$. We use Variational Inference (VI) methods to fit an approximate posterior distribution \citep[see, e.g.,][]{Blei_2017}. VI frames the inference as an optimization problem. The key steps of VI consist of selecting a~parametric family of variational distributions $\mathcal{Q} = \{q_{\bm \phi}, \bm \phi \in \bm \Phi\}$ and determining the parameter $\bm \phi^{*} \in \bm \Phi$ 
minimizing the Kullback-Leibler divergence ($\KL$) of the variational distribution from the true posterior
\begin{align*}
    q_{\bm \phi^{*}}(\bm \theta, \bm \beta^\star, \widetilde{\bm \beta} ) 
    = 
    \underset{q_{\bm \phi} \in \mathcal{Q}}{\argmin} 
    \KL \left(
        q_{\bm \phi}(\bm \theta, \bm \beta^\star, \widetilde{\bm \beta})  
        \,\left\|\, 
        p(\bm \theta, \bm \beta^\star, \widetilde{\bm \beta}  \,|\, \mathbb{Y})
        \right.
    \right).
\end{align*}
This $\KL$ optimization problem is equivalent to maximizing the evidence lower bound ($\ELBO$):
\begin{equation}\label{eq:ELBO}
    \ELBO(\bm \phi) 
    =
    \Eq \left[   
        \log p(\bm \theta, \bm \beta^\star, \widetilde{\bm \beta} ) 
        + 
        \log p(\mathbb{Y} \,|\, \bm \theta, \bm \beta^\star, \widetilde{\bm \beta} ) 
        - 
        \log q_{\bm \phi}(\bm \theta, \bm \beta^\star, \widetilde{\bm \beta} )
    \right]  
\end{equation}
or minimizing the negative $\ELBO$ \citep{Jordan1998, 10.5555/1162264}. Equation~\eqref{eq:ELBO} sums the expectation of the log-likelihood and the log-prior and the entropy of the variational family. 

In the mean-field approach, the variational family $q_{\bm \phi}$ factorizes over its latent variables by considering these variables to be independent and each being governed by their own distribution, i.e., 
\begin{align*}
q_{\bm \phi}(\bm \theta, \bm \beta^\star, \widetilde{\bm \beta} ) 
&= 
\prod_{d,k = 1}^{D,K} q(\theta_{dk}) 
\prod_{k,v = 1}^{K, V} q(\beta^\star_{kv}) 
\prod_{k,v \in \mathcal{S}} q(\widetilde{\beta}_{kv}).
\end{align*}
Only distributions with support on the positive reals are suitable as variational distributions for document-topic and topic-term intensities. We propose to use gamma distributions, matching the prior distributions outlined in Equations~\eqref{eq:prior_betas} and~\eqref{eq:prior_theta}. We denote shape and rate parameters with the superscript `$\shp$' and `$\rte$', respectively. Moreover, we also employ a scaling by document length $N_d = \sum\limits_{v=1}^V y_{dv}$ for parameters $\theta_{dk}$. Including the document length $N_d$ in this way provided empirically a more stable and quicker model fit and induced a superior classification performance. Altogether, we posit as variational distributions 
\begin{align*}
    q(\theta_{dk}) &= \Gamma\left(
        \phi^{\shp}_{\theta_{dk}}, N_d \cdot \phi^{\rte}_{\theta_{dk}}
    \right),
    &
    q(\beta^\star_{kv}) &= \Gamma\left(
        \phi^{\shp}_{\beta^\star_{kv}}, \phi^{\rte}_{\beta^\star_{kv}}
    \right),
    &
    q(\widetilde{\beta}_{kv}) &= \Gamma\left(
        \phi^{\shp}_{\widetilde{\beta}_{kv}}, \phi^{\rte}_{\widetilde{\beta}_{kv}}
    \right).
\end{align*}
Hence, we optimize the $\ELBO$ with respect to the set of variational parameters 
$
\bm \phi = \{
    \bm \phi^{\shp}_{\theta}, 
    \bm \phi^{\rte}_{\theta}$, 
    $\bm \phi^{\shp}_{\beta^\star}, 
    \bm \phi^{\rte}_{\beta^\star}, 
    \bm \phi^{\shp}_{\widetilde{\beta}}, 
    \bm \phi^{\rte}_{\widetilde{\beta}}
\}
\in \bm \Phi = 
\Rpos^{2DK} \times \Rpos^{2KV} \times \Rpos^{2|\mathcal{S}|} 
$.

We use Black Box Variational Inference (BBVI) with stochastic optimization 
and follow \cite{ranganath2013black} to form noisy unbiased gradient estimates of the $\ELBO$ with $S$ Monte Carlo samples from the variational distribution, 
\begin{equation}\label{eq:grad_elbo}
    \nabla_{\bm \phi} \ELBO(\bm \phi) 
    \approx 
    \frac{1}{S}\sum_{s = 1}^{S} 
    \nabla_{\bm \phi} 
    \log q_{\bm \phi}(\bm \theta_s, \bm \beta^\star_s, \widetilde{\bm\beta}_s) 
    \left(
        \log p(\bm \theta_s, \bm \beta^\star_s, \widetilde{\bm\beta}_s, \mathbb{Y}) 
        - 
        \log q_{\bm \phi}(\bm{\theta}_s, \bm{\beta^\star}_s, \widetilde{\bm \beta}_s)
    \right), 
\end{equation}
where $\bm{\theta}_s, \bm{\beta^\star}_s, \widetilde{\bm\beta}_s \sim q_{\bm \phi}(\bm \theta, \bm \beta^\star, \widetilde{\bm \beta} )$ are independent samples from the variational distributions.
These gradient estimates are used to optimize the $\ELBO$ while the updates $\bm \phi$ are determined by the Adam algorithm \citep{kingma2017adam}. Reverse-mode automatic differentiation is used to track all sequences of operations and to compute the gradients during the optimization procedure \citep[see][]{kucukelbir2016automatic}. The whole procedure is shown in Algorithm~\ref{alg:SPF} for $S=1$ which is the value for $S$ we use in our implementation. 
When applying Algorithm~\ref{alg:SPF}, we initialize the variational parameter with $\bm{\phi}^{\shp}_{\theta} = 1$, $\bm{\phi}^{\rte}_{\theta} = \frac{D}{1000}$, $\bm{\phi}^{\shp}_{\beta^\star} = 1$, $\bm{\phi}^{\rte}_{\beta^\star} = \frac{2D}{1000}$, $\bm{\phi}^{\shp}_{\widetilde{\beta}} = \bm{\phi}^{\rte}_{\widetilde{\beta}} = 1$.

\SetKwInput{KwData}{Input}
\SetKwInput{KwResult}{Output}
\SetKwComment{Comment}{> }{ }
\begin{algorithm}[H]
\DontPrintSemicolon
\caption{Seeded Poisson factorization algorithm for $S = 1$}\label{alg:SPF}
\KwData{DTM $\mathbb{Y}$, number of topics $K$, sets of seed words $\mathcal{V}_k$, $k=1,\ldots,K$;\\
\phantom{\textbf{Input: }}prior parameters $a,b,c,d,e,f$, 
initial variational parameter $\bm \phi$;\\
\phantom{\textbf{Input: }}number of epochs $E$,  batch size $|\mathcal{B}|$, learning rate $\rho$.
}
\KwResult{
The last value $\bm{\hat{\phi}}$ when optimizing $\ELBO (\bm \phi)$.
}

\For{epoch $e = 1,2, \ldots , E$}{
    Divide $D$ documents randomly into $B$ batches $\mathcal{B}_b$, $b = 1, \ldots, B$, 
    $|\mathcal{B}_b| \approx |\mathcal{B}|$.

    \For{b in $1:B$}{
    \For{each topic $k \in \{ 1, \ldots, K\}$ and each word $v \in \{1, \ldots, V \}$}{
    Sample $\beta^\star_{kv} \sim \Gamma(\phi^{\shp}_{\beta^\star_{kv}}, \phi^{\rte}_{\beta^\star_{kv}})$.\;
    \eIf{$v \in \mathcal{V}_k$}{
    Sample $\widetilde{\beta}_{kv} \sim \Gamma(\phi^{\shp}_{\widetilde{\beta}_{kv}}, \phi^{\rte}_{\widetilde{\beta}_{kv}})$.
    }{
    Set $\widetilde{\beta}_{kv} = 0$.}
    }
    Compute $\bm{\beta} = \bm{\beta^\star} + \bm{\widetilde{\beta}}$.\;

    \For{each document $d$ in batch $\mathcal{B}_b$}{
        Sample $\theta_{dk} \sim \Gamma(\phi^{\shp}_{\theta_{dk}}, N_d \cdot \phi^{\rte}_{\theta_{dk}})$.\;
        
        \For{$v \in \{1, \ldots, V\}$}{
            Set $\lambda_{dv} = \sum\limits_{k=1}^K \theta_{dk} \beta_{kv}$. \;
            Compute $\log p(y_{dv} \,|\, \bm \theta, \bm \beta^\star, \widetilde{\bm \beta} ) = \log \Pois(y_{dv}\,|\,\lambda_{dv})$. \Comment*[r]{Log-likelihood}
        }
    }
    Set $\log p(\mathbb{Y} \,|\, \bm \theta, \bm \beta^\star, \widetilde{\bm \beta} ) = \frac{D}{|\mathcal{B}_b|} \sum\limits_{d \in \mathcal{B}_b} \sum\limits_{v=1}^V \log p(y_{dv} \,|\, \bm \theta, \bm \beta^\star, \widetilde{\bm \beta} )$.
    \Comment*[r]{Reconstruction}    
    
    Compute $\log p(\bm \theta, \bm \beta^\star, \widetilde{\bm \beta} )$ and $\log q_{\bm \phi}(\bm \theta, \bm \beta^\star, \widetilde{\bm \beta} )$. \Comment*[r]{Prior and entropy}
    Compute $
    \ELBO(\bm \phi) 
    = 
    \log p(\mathbb{Y} \,|\, \bm \theta, \bm \beta^\star, \widetilde{\bm \beta} )  
    + 
    \log p(\bm \theta, \bm \beta^\star, \widetilde{\bm \beta} )
    - 
    \log q_{\bm \phi}(\bm \theta, \bm \beta^\star, \widetilde{\bm \beta} )$.\;
    Compute gradients $\nabla_\phi \ELBO(\bm \phi)$ using automatic differentiation as in Equation~\eqref{eq:grad_elbo}.\;
    Update variational parameter $\bm{\phi}$ with Adam and learning rate $\rho$.

    }
}
\end{algorithm}

\subsection{Post-processing final inference}

After running the model for a sufficient number of epochs $E$, the final value obtained from the VI optimization $\hat{\bm \phi}$ represents the estimate of the variational parameter. 
We summarize the results by determining point estimates for the parameters of interest based on posterior means derived from the posterior approximations through the variational family. In particular, the posterior mean estimates $\hat{\bm \theta}, \hat{\bm \beta}^\star, \hat{\widetilde{\bm \beta}}$ are obtained by determining the means induced by the variational Gamma distributions. In case of document-topic intensities, we obtain posterior means using
\begin{align}\label{eq:mean_topic_intensity}
    \hat{\theta}_{dk} 
    = \hat{\phi}_{\theta_{dk}}^{\shp} / \hat{\phi}_{\theta_{dk}}^{\rte}.
\end{align}
To estimate the topic-term intensities for a topic $k$, we use
\begin{equation} \label{eq:mean_term_intensity}
    \hat{\beta}_{kv}
    =
    \hat{\beta}^\star_{kv} + \hat{\widetilde{\beta}}_{kv}
    = 
    \hat{\phi}_{\beta^\star_{kv}}^{\shp} / \hat{\phi}_{\beta^\star_{kv}}^{\rte}
    +
    \left\{ 
        \begin{aligned}
            &\hat{\phi}_{\widetilde{\beta}_{kv}}^{\shp} / \hat{\phi}_{\widetilde{\beta}_{kv}}^{\rte} && \text{if}\; (k,v) \in \mathcal{S}, \\
            &0 && \text{otherwise.}
        \end{aligned}
    \right.
\end{equation}

The final topic assignment is based on a Naive Bayes classifier \citep[see, e.g.,][]{naivebayes}, i.e., the document is assigned to the topic where the per-document posterior mean estimate is maximal.

We employ standard measures used in classification to evaluate the predictive performance of the SPF topic model when used for automatic text classification with known categories. In particular, we use the following metrics separately for each topic: precision (i.e., correctly assigned documents among all documents assigned to the category), recall (i.e., correctly assigned documents among all documents belonging to the category) and F1-score (harmonic mean of precision and recall). In addition, we obtain aggregate measures using either equal category weights (macro avg) or taking the empirical category frequencies into account (weighted avg). We also provide information on assignment certainty, presenting insights into the model's confidence in its predictions per category, by determining the proportion the document-topic intensity has for the topic the document is assigned to compared to all document-topic intensities. Specifically, we determine the average assignment certainty for true positive (ACTP) and false positive (ACFP) predictions for each category.

Although the primary objective of the SPF topic model is to enhance classification performance, we additionally evaluate topic quality by assessing both topic coherence and topic diversity. Topic coherence is measured using three standard metrics. First, NPMI (Normalized Pointwise Mutual Information; \citealt{lau2014machine}) quantifies the semantic coherence among the top-ranked words of each topic, with higher values indicating better coherence. Second, the UMass score \citep{mimno2011optimizing} relies on document co-occurrence statistics, where fewer negative values suggest greater topic quality. Third, $C_V$ \citep{roder2015exploring} combines several coherence signals using a sliding window, normalized pointwise mutual information, and cosine similarity between word vectors. This metric has been shown to correlate well with human judgment of topic quality. To assess topic diversity, we follow \citet{dieng2019topic} and compute the proportion of unique terms among the top words across all topics, capturing the distinctiveness of the learned topic representations. 

\color{black}

\subsection{Computational details}\label{sec:computational_details}

The SPF model is implemented in Python 3.10. It allows Graphics Processing Unit (GPU) support due to its implementation in the TensorFlow environment. The model's source code is mainly based on TensorFlow 2.18 as well as TensorFlow's add-on library for probabilistic reasoning, TensorFlow Probability 0.25.0. In its standard implementation, SPF uses a batch size of $1\,024$ documents, a learning rate of $0.1$ and trains the model for $150$ epochs. Leveraging TensorFlow's computational graph and gradient tape functionalities, the implementation enables efficient tracking of operations for automatic differentiation during model training. Tokenization and the construction of the document-term matrix (DTM) are carried out using the Python library \textbf{scikit-learn} \citep{scikit-learn}.
The topic coherence and diversity metrics are computed using the Python library \textbf{gensim}  \citep{rehurek_lrec}, based on the default settings for each metric as provided by the library.

The results presented in this paper are compiled locally on a machine with CPU: Intel i5 13600k; GPU: Nvidia RTX 3090; RAM: 32GB DDR5 5600 MHz. Additionally, we employed SPF in an AWS cloud computing environment using an ml.g5dn.xlarge instance\footnote{See instance types: \url{https://docs.aws.amazon.com/sagemaker/latest/dg/notebooks-available-instance-types.html}.} with enhanced GPU support to ensure that the software provided is ready to use in different environments. Our implementation is available as open source software via GitHub.\footnote{See \url{https://github.com/BPro2410/Seeded-Poisson-Factorization}.}

\section{Empirical results}\label{sec:empirical}

We demonstrate the use of SPF on the Amazon Reviews dataset \citep{Kashnitsky_2020} and four benchmark datasets consisting of text corpora of varying size and text classifications with a varying number of categories. We provide a detailed analysis of the application on the Amazon Reviews dataset to illustrate topic assessment as well as compare SPF to other LDA-based seeded topic models with respect to classification performance, topic coherence and diversity as well as computational efficiency. In addition, we perform a robustness analysis to assess the impact of the hyperparameters. The application to the four benchmark datasets indicate the general applicability of SPF for automatic text classification and the computational efficiency.

When fitting the SPF model in these empirical applications, we set the number of topics to correspond to the number of categories and follow the lines of \citet{watanabe} to construct a balanced lexicon of ten frequently occurring seed words for each category corresponding to a topic. To generate these seed words, we compute the TF-IDF (Term Frequency-Inverse Document Frequency) values for each word within each category. We select the top-10 words from the TF-IDF matrix as the seed words for each topic. This process is conducted in an automatic way to minimize subjectivity and to ensure that the seed word selection process remains as objective as possible.

\color{black}

\subsection{Application to the Amazon Reviews dataset}

The Amazon Reviews dataset consists of customer feedback on products sold through Amazon from the following six level-1 product categories: health personal care, toys games, beauty, pet supplies, baby products, and grocery gourmet food. Each observation consists of the review text and the information on the product category. 
To prepare the text data, we apply the following pre-processing steps:  text normalization (conversion to lowercase), removal of stop words, and exclusion of words appearing fewer than two times in the corpus. We construct the DTM using a sample of $30\,000$ documents, each containing at least seven words. The resulting matrix $\mathbb{Y}$ includes $D = 30\,000$ non-empty customer feedback entries and a vocabulary size of $V = 23\,135$ unique terms. On average, the documents in $\mathbb{Y}$ contain 42.3 words, with $5$\% and $95$\% quantiles of $15.0$ and $100.0$, respectively. Table~\ref{tab:data-set-characteristics} presents a summary of the final sample of documents and the pre-specified seed words per product category selected using the proposed automatic procedure.

\begin{table}[t!]
  \begin{adjustbox}{max width=\textwidth}
  \centering
  \begin{tabular}{l l c l}
    \toprule
    \textbf{Product category} & \textbf{Topic} &
    \textbf{Count}      & \textbf{Seed words} \\
    \midrule
      Toys games & Toys               & 8\,092        & toy, game, play, fun, old, son, year, loves, kids, daughter\\
      Health personal care & Health  & 6\,938        &  product, like, razor, shave, time, day, shaver, better, work, years\\ 
      Beauty products & Beauty                & 4\,072         &  hair, skin, product, color, scent, smell, used, dry, using, products\\
      Baby products & Baby         & 4\,635       &  baby, seat, diaper, diapers, stroller, bottles, son, pump, gate, months\\
      Pet supplies & Pets           & 3\,792         &  dog, cat, litter, cats, dogs, food, box, collar, water, pet\\
      Grocery gourmet food & Grocery    & 2\,471       &  tea, taste, flavor, coffee, sauce, chocolate, sugar, eat, sweet, delicious\\
    \bottomrule
  \end{tabular}
  \end{adjustbox}
  \caption{Overview of the final sample: document counts and seed words by product category.}
  \label{tab:data-set-characteristics}
\end{table}

\subsubsection{Topic assessment}

We examine the topics inferred by SPF based on the approximate posterior mean topic-term intensities $\hat{\bm{\beta}}_k$ for each topic $k$. Table~\ref{tab:betas} presents the top-14 terms with the highest approximate mean intensities per topic, after removing stop-word-like terms that provide no contextual information. Bold terms represent seed words. The mean term intensity (provided in parentheses) is calculated as the sum of the seeded term intensity and the unseeded term intensity, as defined in Equation~\eqref{eq:mean_term_intensity}. These high-intensity words per topic enable the characterization of the topic as well as the assessment of how influential the seed words were. 

Clearly the pre-defined seed words exhibit a strong presence among the most pertinent terms for all topics. However, the number of seed words included in the top-14 words with highest intensity varies across topics. For topic `Toys`, all 10 seed words are also included in the list of 14 most pertinent terms for this topic. This number decreases to six for `Health', eight for `Beauty', seven for `Baby' and eight for `Pets'. The lowest number of seed words are included in the list of 14 most pertinent terms for the topic `Grocery' where only four out of the ten seed words are listed. 

Table~\ref{tab:betas} reveals that the model effectively prioritizes not only the explicitly defined seed words but also identifies and assigns significant weight to relevant additional terms that are not prespecified as seed words. For example, for topic `Health' the seed word `day' was specified and also `days' is included among the 14 most pertinent terms. For topic `Baby' not only the seed word `son' is included but also `daughter'. Inspecting the `Baby' topic further by also assessing additional terms with high intensity indicates that SPF did not only assign high relevance to expected seed words such as `seat' (14.88) and `son' (12.07), but also recognized terms like `bed' (6.71) and `sleep' (6.31) as highly pertinent to the topic. These terms align with the product subcategory `sleep positioners', which falls under the broader `Baby' category, demonstrating the model's nuanced understanding of topic content and its ability to discern contextually important terms. 

The topic `Grocery' fails to capture most of the seed words among the 14 most pertinent terms. However, the terms with high intensity suggest that this topic captured an additional aspect in customer reviews which relates not to the product but to the purchase process. For example, terms like `store' ($7.34$) and `shipping' ($7.20$) have a high prevalence in the `Grocery' topic.

\begin{table}[t!]
\begin{adjustbox}{max width=\textwidth}
\begin{tabular}{@{}l|l|l|l|l|l@{}}
\toprule
\textbf{Toys} & \textbf{Health} & \textbf{Beauty} & \textbf{Baby} & \textbf{Pets} & \textbf{Grocery} \\ \midrule
\textbf{toy (39.67)}  & \textbf{product} (25.40)  & \textbf{hair (48.66)}  & \textbf{baby (32.00)}  & \textbf{dog (24.57)}  & amazon (19.30)  \\
\textbf{old (30.27)}  & \textbf{time (17.00)}  & \textbf{product (30.67)}  & use (20.89)  & \textbf{water (16.68)}  & like (18.62)  \\
\textbf{game (22.08)}  & \textbf{work (14.58)}  & like (24.75)  & \textbf{seat (14.88)}  & \textbf{cat (15.62)}  & product (16.65)  \\
\textbf{play (21.54)}  & \textbf{years (13.01)}  & use (22.91)  & easy (14.60)  & \textbf{box (14.19)}  & \textbf{taste (13.70)}  \\
\textbf{year (20.48)}  & \textbf{day} (12.70)  & \textbf{skin (22.41)}  & little (12.92)  & product (12.05)  & \textbf{tea (13.16)}  \\
\textbf{fun (19.22)}  & used (11.64)  & really (13.63)  & \textbf{son (12.07)}  & \textbf{dogs (10.55)}  & price (10.24)  \\
\textbf{loves (18.70)}  & good (9.92)  & \textbf{color (12.06)}  & old (11.57)  & \textbf{cats (10.53)}  & \textbf{flavor (9.80)}  \\ 
great (18.23)  & works (7.48)  & \textbf{smell (11.93)}  & \textbf{months (11.25)}  & \textbf{litter (9.94)}  & buy (7.66)  \\
like (17.62)  & days (7.24)  & \textbf{dry (10.64)}  & fit (10.13)  & small (9.31)  & store (7.34)  \\
little (17.09)  & batteries (7.19)  & time (10.22)  & car (9.92)  & time (8.78)  & shipping (7.20)  \\
\textbf{son (16.68)}  & battery (6.55)  & good (9.90)  & daughter (9.08)  & little (8.29)  & order (7.10)  \\
\textbf{daughter (14.85)}  & pain (6.43)  & \textbf{products (8.83)}  & \textbf{diaper (8.01)}  & plastic (8.05)  & \textbf{eat (6.02)}  \\
bought (12.56)  & \textbf{razor (6.29)}  & face (8.79)  & \textbf{bottles (7.24)}  & clean (7.46)  & food (6.00)  \\
\textbf{kids (12.52)}  & reviews (6.29)  & \textbf{scent (8.44)}  & \textbf{pump (6.97)}  & \textbf{food (7.41)}  & protein (5.44)  \\ \bottomrule
\end{tabular}
\end{adjustbox}
\caption{High-intensity words per topic. Mean intensities are shown in brackets. Bold words are seed words.}
\label{tab:betas}
\end{table}

Table~\ref{tab:betas} also illustrates that SPF is able to assign distinctly different intensities to the seed words as well as other terms with high intensity within their respective topics. E.g., within the `Pets' category, `dog' ($24.57$) plays a dominant role, whereas other seed words, like `food' ($7.41$), display a markedly lower mean intensity. This contrast highlights the ability of SPF to estimate the uneven influence of seed words in defining a topic. This property of the SPF topic model is important to also mitigate the risk of potential misspecifications that may arise due to incomplete domain knowledge during the initial selection of seed words. To empirically assess the influence of misspecified seed words, we conducted an additional analysis evaluating the performance of SPF when an inappropriate seed word is assigned to a topic. Specifically, we fitted the SPF model with `dog' as a seed word for the `Beauty' topic. The results indicate that SPF effectively recognizes that the term contributes minimal to no informational value in this context. In particular, the inferred variational distribution was $\widetilde{\beta}_{beauty, dog} \sim \Gamma(0.25, 5.66)$. These findings underscore SPF's ability to adaptively assign importance to seed words, thereby reducing the impact of initial specification errors.

\subsubsection{Classification performance}

Next, we measure the classification performance of SPF based on approximate posterior means of the document-topic intensities, where each document vector is a $K$-dimensional vector of approximate mean intensities $\hat{\bm{\theta}}_d \in \Rpos^K$, see Equation~\eqref{eq:mean_topic_intensity}. According to the Naive Bayes classifier, the topic with the highest approximate mean intensity in $\hat{\bm{\theta}}_d$ is assigned as the predicted topic for document $d$. We assess the classification performance separately for each category.

\begin{table}[!ht]
    \centering
    \setlength{\tabcolsep}{7pt} 
    \begin{adjustbox}{max width=\textwidth}
    \begin{tabular}{l|ccc|cc} 
        \toprule
        \textbf{Category} & \textbf{Precision} & \textbf{Recall} & \textbf{F1-score} & \textbf{ACTP} & \textbf{ACFP} \\ 
        \midrule
        Toys     &  0.92 &  0.82 &  0.87 &  0.68 &  0.51 \\ 
        Health   &  0.75 &  0.46 &  0.57 &  0.64 &  0.60 \\ 
        Beauty   &  0.68 &  0.79 &  0.73 &  0.71 &  0.54 \\ 
        Baby     &  0.71 &  0.78 &  0.74 &  0.70 &  0.54 \\ 
        Pets     &  0.61 &  0.76 &  0.74 &  0.65 &  0.52 \\ 
        Grocery  &  0.51 &  0.94 &  0.66 &  0.78 &  0.54 \\ 
        \midrule
        Macro avg   &  0.71 &  0.76 &  0.72 &  &  \\ 
        Weighted avg &  0.75 &  0.73 &  0.73 &  & \\ 
        \bottomrule
    \end{tabular}
    \end{adjustbox}
    \caption{Classification performance of the SPF topic model on Amazon customer feedback, including assignment certainty of true positives  (ACTP) and false positives (ACFP). The overall accuracy is 0.73.}
    \label{tab:classification_results}
\end{table}

The classification performance of the SPF topic model is summarized in Table~\ref{tab:classification_results}. Clearly, SPF provides excellent classification performance in categorizing Amazon customer feedback across all six product categories, despite slight differences among categories. The overall accuracy of the model is 0.73, which is consistent with the weighted average F1-score (0.73), accounting for the varying sample sizes across categories. The macro average F1-score (0.72) is slightly lower, reflecting the imbalanced performance among categories. For instance, the highest F1-score is observed for the `Toys' category ($0.87$), reflecting both high precision ($0.92$) and recall ($0.82$). This suggests a strong alignment between the predicted and true labels. Conversely, the `Grocery' category achieves the lowest F1-score ($0.66$), driven by a significant imbalance between precision ($0.51$) and recall ($0.94$). This discrepancy indicates a tendency to over-predict the `Grocery' category, resulting in higher recall at the cost of precision. The over-prediction of the `Grocery' topic is also reflected in the highest ACTP score of $0.78$. By contrast, the `Health' category shows an inverted pattern, with a high precision ($0.75$) but a much lower recall ($0.46$), indicating under-representation in predictions. In the `Health' category, the model exhibits the lowest ACTP score at $0.64$, indicating that SPF is on average less confident in the assignment compared to all other categories when correctly assigning a review. At the same time, the ACFP score is the highest among all categories at $0.60$. This combination of low confidence in true positives and high confidence in false positives highlights the model's particular struggle with distinguishing health-related feedback, emphasizing the need for further refinement in this category.  Table~\ref{tab:betas} shows that the topic-term intensities of seed words for the `Grocery' and `Health' category are in general not as strong as the ones for the other categories. This likely contributes to the lower classification performance observed for these categories. The correct specification of seed words appears to be a crucial factor in achieving high classification performance, as demonstrated by the results in the `Toys' category. This highlights the model's particularly strong performance in categories with distinct linguistic characteristics. However, categories like `Grocery' and `Health' reveal areas where the model might benefit from further refinement, such as enhanced domain-specific seed words or adjustments to address label imbalance. In addition, including an additional unseeded topic that captures feedbacks discussing the purchasing process rather than the product could improve the categorization of the feedback items. 
Nevertheless, the overall accuracy and balanced macro and weighted averages suggest a generally robust model, even if further refinement could enhance performance in underperforming categories.

\subsubsection{Comparison to existing methods and scalability}\label{sec:pc}

To evaluate the classification performance and the computational efficiency of the SPF topic model in comparison with other guided topic models, we also fit KeyATM \citep{eshima2023keyword} and SeededLDA \citep{watanabe} to the Amazon corpus. We complement this with a comparison to the standard PF model.
We investigate in particular how the classification performance and run-times change with the number of documents $D$ in the corpus, varying $D$ from $1k$, over $5k$ and $10k$ to $30k$. The evaluation criteria include the run-time (in minutes) as well as the classification performance metrics accuracy, precision, recall, and F1-score. These computational experiments were conducted using the hardware setup described in Section~\ref{sec:computational_details}.

When fitting the standard PF topic model, the inferred topics are unlabeled and their ordering is arbitrary. Evaluating the classification accuracy requires identifying an alignment between the predicted topics and the true category labels. To do so, we apply the Hungarian algorithm \citep{kuhn1955hungarian} to obtain the permutation of predicted topics that maximizes accuracy. This approach is a form of optimal label permutation commonly used in clustering and unsupervised learning evaluation, and is equivalent to solving the linear sum assignment problem \citep{munkres1957algorithms}. Based on this optimal mapping, the predicted labels are accordingly remapped and the classification metrics calculated.

\color{black}
For all models, we set the number of topics to $K=6$ and limited the input data to customer feedback only, excluding any additional metadata. Each model was trained using the same set of seed words and the default values for model and prior specifications suggested / implemented in the software packages. Using the same number of MCMC iterations for KeyATM and SeededLDA as well as the number of epochs for model fitting regardless of the number of documents $D$ led to poor predictive performance results in the case where only very few documents were included in the corpus. We thus increased the number of MCMC iterations / number of epochs for $D$ equal to $5k$ and $1k$. In particular, we used 1\,500 MCMC iterations and 150 epochs for $D \in \{10k, 30k\}$ and doubled this number to 3\,000 MCMC iterations and 300 epochs for $D = 5k$ and tripled the number to 4\,500 MCMC iterations and 450 epochs for $D = 1k$. 

Table~\ref{tab:model_fit} provides the results for this comparison. The run-time comparison clearly shows that regardless of corpus size, SPF 
has a comparable run-time to PF and it 
always has the shortest run-times compared to both SeededLDA and KeyATM.
The difference in run-time 
increases with the corpus size. While for a corpus of size $1k$ documents, the run-times of KeyATM and SeededLDA are only approximately 1.5 times the run-time of SPF, the run-times increase by a factor of 3 to 5 times for a corpus of size $30k$ documents. 

\begin{table}[!t]
\centering
\begin{adjustbox}{max width=0.95\textwidth}
\begin{tabular}{@{}lrrrrlrrrr@{}}
\toprule
& \multicolumn{4}{c}{$30k$ documents}                                           & & \multicolumn{4}{c}{$10k$ documents}                                            \\ \cmidrule(lr){2-5} \cmidrule(l){7-10} 
& SPF   & KeyATM & SeededLDA & PF        &   & SPF   & KeyATM & SeededLDA & PF        \\
\textit{Time (min:sec)} & \textit{1:07} & \textit{5:27} & \textit{3:35} & \textit{1:04}& & \textit{0:19} & \textit{1:41} & \textit{1:04} & 0:19\\ \midrule
Accuracy   & \textbf{0.73}  & \textbf{0.73}   & 0.65      & 0.55&   & \textbf{0.72}  & 0.57   & \underline{0.63}      & 0.50\\
Precision  & \underline{0.71}  & \textbf{0.73}   & 0.63      & 0.52&   & \underline{0.71}  & \textbf{0.72}   & 0.62      & 0.47\\
Recall     & \textbf{0.76}  & \underline{0.72}   & 0.68      & 0.56&   & \textbf{0.75}  & 0.52   & \underline{0.66}      & 0.51\\
F1-score   & \textbf{0.72}  & \textbf{0.72}   & 0.65      & 0.53&   & \textbf{0.71}  & 0.55   & \underline{0.63}      & 0.48\\ \midrule
&         &        &           &           &   &        &        &           &           \\ \midrule
& \multicolumn{4}{c}{$5k$ documents}                                            & & \multicolumn{4}{c}{$1k$ documents}                                            \\ \cmidrule(lr){2-5} \cmidrule(l){7-10} 
& SPF   & KeyATM & SeededLDA & PF        &   & SPF   & KeyATM & SeededLDA & PF        \\
\textit{Time (min:sec)} & \textit{0:09} & \textit{0:44} & \textit{0:34} & \textit{0:09}& & \textit{0:06} & \textit{0:09} & \textit{0:09} & \textit{0:07}\\
\midrule
Accuracy   & \textbf{0.70}  & 0.44   & \underline{0.62}      & 0.61&   & \textbf{0.63}  & 0.29   & \underline{0.58}      & 0.32\\
Precision  & \underline{0.68}  & \textbf{0.70}   & 0.61      & 0.59&   & \textbf{0.61}  & 0.50   & \underline{0.57}      & 0.33\\
Recall     & \textbf{0.73}  & 0.38   &  \underline{0.64}      & \underline{0.64}&   & \textbf{0.67}  & 0.21   & \underline{0.59}      & 0.35\\
F1-score   & \textbf{0.69}  & 0.39   & \underline{0.62}      & 0.60&   & \textbf{0.62}  & 0.15   & \underline{0.57}      & 0.32\\
\bottomrule
\end{tabular}
\end{adjustbox}
\caption{Classification performance across models and corpus sizes. \textbf{Bold:} The highest score. \underline{Underline:} The second highest score.}
\label{tab:model_fit}
\end{table}

In terms of accuracy, SPF demonstrates superior or on par performance compared to the other methods across all corpora sizes. The unguided PF topic model generally exhibits the weakest predictive performance among the evaluated methods. This is expected due to its completely unsupervised nature, which lacks any form of guidance during training. In contrast, the SPF model -- despite incorporating only a minimal amount of domain knowledge -- achieves substantially better predictive results, highlighting the effectiveness of even weak supervision in improving classification performance with topic models when categories can be pre-specified. For the largest corpus ($D = 30k$), SPF achieves an accuracy of $0.73$, equivalent to that of KeyATM and higher than SeededLDA's $0.65$, while PF only has an accuracy of 0.55. As corpus size decreases, SPF maintains a higher accuracy, notably outperforming competing methods for smaller corpora. In particular, SPF achieves high accuracy up to a corpus size of $5k$ with a considerable drop in performance only observed in case $1k$ documents are included in the corpus. Also, SeededLDA maintains a similar -- although at a lower level -- accuracy regardless of corpus size with only a slight reduction in the case of a small corpus size. KeyATM is most severely affected by a decrease in corpus size with the accuracy values dropping from 0.73 to 0.29. SPF also outperforms the other methods with respect to the other predictive performance criteria such as precision, recall and F1-score.

Table~\ref{tab:coherence} compares SPF, KeyATM, SeededLDA, and plain PF in terms of topic coherence and diversity across varying corpus sizes. For each model, the scores are computed based on the top-10 highest-ranked terms per topic. Overall, SPF exhibits competitive performance in the topic coherence metrics. While SeededLDA achieves the highest $C_V$ coherence scores in most settings, SPF consistently matches or outperforms its competitors on NPMI and UMass. For example, at corpus sizes of 30$k$ and 10$k$, SPF achieves UMass scores of $-1.80$ and $-1.75$, respectively -- higher (i.e., better) than those of SeededLDA, and comparable to or slightly below those of KeyATM. At the smallest data size (1$k$ documents), SPF clearly outperforms both baselines on UMass, highlighting its robustness in low-resource scenarios. Interestingly, SPF and PF yield similar results in coherence and diversity across all corpus sizes.

In terms of topic diversity, SPF generally falls between KeyATM and SeededLDA, while performing comparably to PF. It provides more diverse topics than KeyATM on the 30$k$ dataset but remains below SeededLDA, which achieves the highest diversity scores across most corpus sizes. At smaller scales (i.e., 5$k$ and 1$k$ documents), KeyATM surpasses SPF in diversity, although this comes at the cost of lower coherence. These results highlight the design trade-offs between models. We emphasize that while the primary objective of the SPF model is to enhance classification performance, it adeptly balances coherence and diversity, with a consistent advantage in maintaining topic quality under data constraints. Tables~\ref{tab:model_fit} and \ref{tab:coherence} together highlight the importance of not relying solely on topic quality metrics when evaluating models for classification tasks. For instance, in the 5$k$ document setting, KeyATM attains a higher topic diversity score ($0.82$) than SPF ($0.68$). Nevertheless, this increased diversity is accompanied by a markedly lower predictive accuracy -- $0.44$ for KeyATM compared to $0.70$ for SPF -- illustrating that greater topic diversity does not necessarily imply superior classification performance.

\color{black}

\begin{table}[!t]
\centering
\begin{adjustbox}{max width=0.95\textwidth}
\begin{tabular}{@{}lrrrrlrrrr@{}}
\toprule
& \multicolumn{4}{c}{$30k$ documents}                                   & & \multicolumn{4}{c}{$10k$ documents}                                    \\ \cmidrule(lr){2-5} \cmidrule(l){7-10} 
& SPF & KeyATM & SeededLDA & PF       &   & SPF & KeyATM & SeededLDA & PF       \\
\midrule
NPMI       & \underline{0.15} & \textbf{0.16}   & 0.14      & \underline{0.15}&   & \underline{0.11} & \textbf{0.12}   & \underline{0.11}      & 0.10
\\
UMass      & $-$1.80 & \textbf{$-$1.68} & $-$2.03   & \underline{$-$1.78}&   & \underline{$-$1.75} & $-$2.02 & $-$2.22   & \textbf{$-$1.74}
\\
$C_V$      & 0.58 & \underline{0.59}   & \textbf{0.64}      & 0.57&   & 0.58 & \underline{0.61}   & \textbf{0.64}      & 0.56
\\
Diversity  & \underline{0.67} & 0.57   & \textbf{0.83}     & \underline{0.67} &   & 0.63 & \underline{0.73}   & \textbf{0.83}      & 0.63\\
\midrule
&      &        &           &          &   &      &        &           &          \\ \midrule
& \multicolumn{4}{c}{$5k$ documents}                                    & & \multicolumn{4}{c}{$1k$ documents}                                  \\ \cmidrule(lr){2-5} \cmidrule(l){7-10} 
& SPF & KeyATM & SeededLDA & PF       &   & SPF & KeyATM & SeededLDA & PF       \\
\midrule
NPMI       & \underline{0.13} & 0.11   & 0.08      & \textbf{0.14}&   & \underline{0.07} & 0.02   & 0.02      & \textbf{0.11}\\
UMass      & \textbf{$-$1.81} & $-$2.22 & $-$2.57   & \underline{$-$1.88}&   & \textbf{$-$1.71} & $-$4.63 & $-$3.36   & \underline{$-$1.89}\\
$C_V$      & 0.54 & \underline{0.57}   & \textbf{0.59}      & 0.52&   & \textbf{0.54} & \underline{0.52}   & \underline{0.52}      & 0.51\\
Diversity  & 0.68 & \underline{0.82}   & \textbf{0.87}      & 0.68&   & 0.63 & \textbf{0.97}   & \underline{0.92}      & 0.67\\
\bottomrule
\end{tabular}
\end{adjustbox}
\caption{Coherence and diversity comparison across different corpora sizes. \textbf{Bold:} The highest score. \underline{Underline:} The second highest score.}
\label{tab:coherence}
\end{table}

\begin{figure}[!t]
    \centering
        \includegraphics[width=0.8\textwidth]{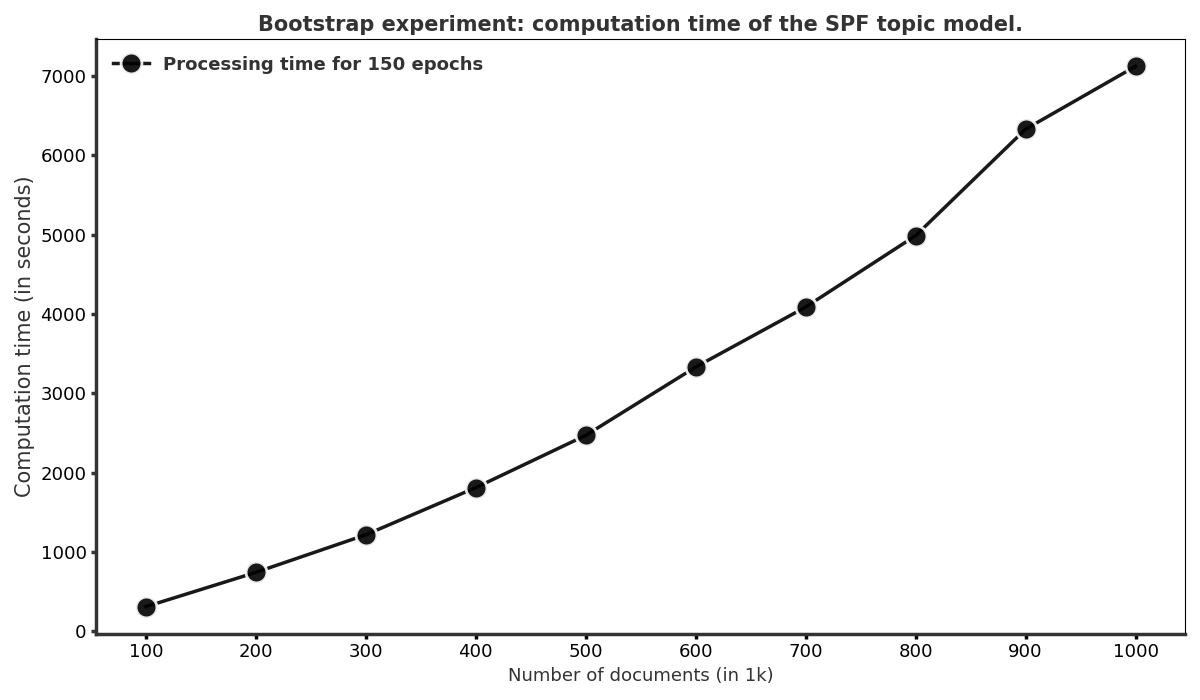}
    \caption{Processing time for the bootstrap experiment. \label{fig:bootstrap_runtime}}
\end{figure}

To evaluate the scalability of SPF, we conduct a bootstrap experiment where we draw documents with replacement from the available corpus to obtain corpora of different size. In particular, we increase the number of documents $D$ in increments of $100k$, up to a total of $D = 1\,000\,000$ documents. This bootstrap experiment was conducted exclusively with the SPF model, as both KeyATM and SeededLDA were unable to handle such large corpora on the hardware described in Section~\ref{sec:computational_details}. Figure~\ref{fig:bootstrap_runtime} visualizes the run-times observed, indicating a roughly linear increase in run-times as the corpus size grows. This increase may be attributed to the increase in the model's local variational parameters $\phi^\rte_\theta$ and $\phi^\shp_\theta$ with the number of documents. Figure~\ref{fig:bootstrap_runtime} shows that SPF successfully processes the corpus with $1\,000\,000$ documents in approximately 2 hours, demonstrating its ability to handle large-scale datasets efficiently.

\subsubsection{Robustness checks}

We systematically vary key parameters of the specification and estimation of the SPF topic model to evaluate how the performance changes under different conditions. Table~\ref{tab:excel_data} presents the resulting 13 scenarios and the corresponding classification performance results obtained when fitting SPF in these scenarios. 

\begin{table}[!t]
\centering
\begin{adjustbox}{max width=\textwidth}
\begin{tabular}{l|c|cccccccccccc}
\toprule
 & (1) & (2) & (3) & (4) & (5) & (6) & (7) & (8) & (9) & (10) & (11) & (12) & (13) \\
\midrule
\textbf{Setting}& & & & & & & & & & & & &\\
Number of documents & 30k & 30k & 30k & 30k & 30k & 30k & 30k & 30k & 30k & 30k & 30k & \cellcolor{lightgray} 1k& \cellcolor{lightgray} 1k\\
Seeded topics & 6 & 6 & 6 & 6 & 6 & 6 & 6 & \cellcolor{lightgray} 5& 6 & 6 & 6 & 6 & 6 \\
Unseeded topics & 0 & 0 & 0 & 0 & 0 & 0 & 0 & \cellcolor{lightgray} 1& \cellcolor{lightgray} 1& 0 & 0 & 0 & 0 \\
$\widetilde{\beta}$ shape prior parameter & 1.0 & 1.0 & 1.0 & 1.0 & 1.0 & 1.0 & 1.0 & 1.0 & 1.0 & \cellcolor{lightgray} 0.3& \cellcolor{lightgray} 2.0& \cellcolor{lightgray} 0.3& \cellcolor{lightgray} 2.0\\
Seed words per topic & 10 & \cellcolor{lightgray} 5 & 10 & 10 & 10 & 10 & 10 & 10 & 10 & 10 & 10 & 10 & 10 \\
Batch size & 1\,024 & 1\,024 & 1\,024 & 1\,024 & \cellcolor{lightgray} 512& \cellcolor{lightgray} 512& \cellcolor{lightgray} 512& 1\,024 & 1\,024 & 1\,024 & 1\,024 & 1\,024 & 1\,024 \\
Learning rate & 0.1 & 0.1 & \cellcolor{lightgray} 0.01& \cellcolor{lightgray} 0.01& 0.1 & \cellcolor{lightgray} 0.01& \cellcolor{lightgray} 0.01& 0.1 & 0.1 & 0.1 & 0.1 & 0.1 & 0.1 \\
Epochs & 150 & 150 & 150 & \cellcolor{lightgray} 300& 150 & 150& \cellcolor{lightgray} 300& 150 & 150 & 150 & 150 & \cellcolor{lightgray} 300 & \cellcolor{lightgray} 300 \\
\hline
\textbf{Performance metrics}& & & & & & & & & & & & &\\
Accuracy & 0.73 & 0.70 & 0.50 & 0.66 & 0.73 & 0.64 & 0.71 & 0.73& 0.67 & 0.73 & 0.73 & 0.62 & 0.64 \\
Precision & 0.71 & 0.68 & 0.58 & 0.65 & 0.71 & 0.64 & 0.69 & 0.71& 0.65 & 0.71 & 0.72 & 0.60 & 0.62 \\
Recall & 0.76 & 0.74 & 0.55 & 0.70 & 0.76 & 0.68 & 0.74 & 0.76& 0.60 & 0.76 & 0.76 & 0.65 & 0.68 \\
F1-score & 0.72 & 0.69 & 0.51 & 0.65 & 0.72 & 0.63 & 0.69 & 0.72& 0.61 & 0.72 & 0.72 & 0.61 & 0.63 \\
\bottomrule
\end{tabular}
\end{adjustbox}
\caption{Robustness checks. Analysis of 13 different scenarios with varying settings regarding the data, model specification and estimation. The setting for the benchmark model is shown in Scenario (1). Changes in the settings, compared to the base scenario, are indicated by a gray background.}
\label{tab:excel_data}
\end{table}

\paragraph{Effect of the number of seed words.} We examine the impact of reducing domain knowledge on classification performance by reducing the number of seed words per topic from 10 to 5 (Scenario 2). As expected, decreasing the amount of seed words results in lower predictive performance. However, halving the number of seed words did result only in a slight decrease in predictive performance, underscoring the importance of being able to select at least a few meaningful seed words to characterize topics in order to achieve superior results using the seeded approach when fitting topic models.  

\paragraph{Effect of learning rate, epochs and batch size.}
In Scenarios 3, 4, 6 and 7 the learning rate is reduced to 0.01 from 0.1 in the base scenario. The results indicate that in particular lowering the learning rate increases the number of epochs required for the negative \(\ELBO\) to converge, suggesting slower optimization (see Scenario 4). 

For the scenarios considered, increasing the number of epochs (Scenarios 4, 7, 12, 13) does not improve the performance. However, a higher number of epochs was in particular used when the simultaneous change of another setting induced the need for more epochs, such as the reduction of the learning rate (Scenarios 4 and 7) or a lower number of documents (Scenarios 12 and 13). 

Reducing the batch size (Scenarios 5, 6, 7) seems to have hardly any impact. Similar results are obtained in particular for Scenario 5 where this is the only setting change. Scenario 7 suggests that lowering the learning rate, reducing the batch size and increasing the number of epochs yields good results, highlighting the interplay between these hyperparameters. 

\paragraph{Effect of varying the number of topics $K$.}
We also examine the effect of altering the number of topics $K$ in two different ways. First, we remove the a-priori information for the `Grocery' topic, estimating the model with five seeded topics and one unseeded topic (Scenario 8).  Second, we add an additional unseeded topic to the six seeded product categories (Scenario 9).  

Dropping the seed words for one topic but otherwise keeping the number of topics (Scenario 8) results in the same good performance as in the base scenario. In the case where the `Grocery' topic is excluded and an unseeded topic is added, SPF allocates $4\,826$ customer reviews to the unseeded category, accurately identifying $2\,331$ out of $2\,471$ instances as belonging to the `Grocery' category. Inspecting the words with the highest topic-term intensities for the unseeded topic indicates that this topic effectively captures the `Grocery' topic. High-intensity words emerging in this setting are `taste' (12.84), `tea' (12.13), and `flavor' (8.78). 

In Scenario 9, adding an additional unseeded topic leads to reduced model performance ($0.67$ compared to $0.73$ in the baseline scenario), as the fixed number of six product categories means that assigning a customer review to the unseeded topic constitutes a misclassification in this context. However, an analysis of topic-term distributions for the unseeded topic reveals that SPF assigns reviews to the unseeded topic when customers primarily discuss the purchasing process rather than specific product characteristics. To give an example, terms such as `time', `shipping', `store', `order' and `online' exhibit high intensities within the unseeded topic. SPF was therefore capable to identify the additional latent topic present in customer reviews which relates to purchasing and delivery experiences, which -- while not tied to specific product categories -- is also of significant business relevance. Overall these findings illustrate SPF's robust capability to discern meaningful latent topics even in the absence of comprehensive domain-specific seeding.

\paragraph{Effect of the shape parameter $c$ on $\widetilde{\beta}$.}
We explored the impact of varying the a-priori relevance assigned to seed words by adjusting the shape parameter $c$ of the prior of the seeded topic-term intensities (Scenarios 10–13). Our findings indicate that the selection of $c$ has minimal impact when the data size is large, i.e., $D = 30\,000$. In this case, the influence of the prior is outweighed by the substantial information in the data (see Scenarios 10 and 11 with an accuracy of 0.73 each). 

Changing the value of the shape parameter $c$ for a smaller dataset ($D = 1\,000$) indicates that this has some effect on model accuracy. In this case, the model accuracy is slightly higher for a more informative prior setting compared to a setting where only a small amount of additional weight is imposed on the seed words, i.e., an accuracy of $0.64$ is obtained in Scenario 13 compared to $0.62$ in Scenario 12. This observation underscores the importance of balancing prior informativeness with dataset size for optimal model performance.

\subsection{Application to four benchmark datasets}

We evaluate SPF model's performance and general applicability using four additional publicly available text corpora. These corpora encompass various domains and in particular also feature a wide range of different number of topics to be estimated, which allows to further validate the SPF approach. Pre-processing involves normalization (conversion to lowercase), tokenization and the removal of documents that result in zero tokens after processing. Dataset statistics are summarized in Table~\ref{tab:exp_data}. 

\begin{itemize}[label=--]
    \item \textbf{Banking} \citep{Casanueva2020}\footnote{\url{https://huggingface.co/datasets/banking77}} is a dataset containing customer service queries from the banking sector. It includes over $10\,000$ labeled queries across 77 fine-grained banking-related intents. We use the query text as the document input and align the pre-defined intent labels with the seed topics. Stop words are retained during pre-processing, resulting in a vocabulary size of $V = 2\,320$.

    \item \textbf{DBPedia} \citep{NIPS2015_250cf8b5}\footnote{\url{https://huggingface.co/datasets/dbpedia_14}} provides structured information extracted from Wiki\-pedia. For our experiments, we use a classified subset consisting of 14 non-overlapping categories (e.g., company, artist, place). The title and abstract fields are concatenated to form the document text, and the 14 class labels are used to derive the seed words. During pre-processing, stop words are removed, and the vocabulary is limited to the top $V = 25\,000$ most frequent terms.

    \item \textbf{20NG} \citep{Lang95}\footnote{\url{https://scikit-learn.org/0.19/datasets/twenty_newsgroups.html}} is a well-known dataset composed of posts to 20 different newsgroups. We treat the post content as document and use the newsgroup information (e.g., \emph{comp.graphics}, \emph{sci.space}) as classes for the seed words. We also remove stop words during pre-processing and limit the vocabulary to the $25\,000$ most frequent terms.

    \item \textbf{Ledgar} \citep{tuggener-etal-2020-ledgar}\footnote{\url{https://aclanthology.org/2020.lrec-1.155/}} is a large-scale multi-label corpus of legal clauses extracted from SEC filings. It contains almost 100\,000 contractual provisions annotated with over 12\,000 clause types. For our experiments, we use the LEX-GLUE LEDGAR subset\footnote{\url{https://huggingface.co/datasets/coastalcph/lex_glue}} with around 100 clause types, treating clause texts as documents and derive the seed words based on the clause types. We remove stop words during pre-processing, which results in a vocabulary of $18\,476$ unique terms.
\end{itemize}

\begin{table}[t!]
\centering
\small
\begin{adjustbox}{max width=\textwidth}
\begin{tabular}{lrrrrlcc cccccccc}
\toprule
\multirow{3}{*}{\textbf{Dataset}} & \multicolumn{4}{c}{\textbf{Characteristics}} & \multirow{3}{*}{\textbf{Model}} & \multicolumn{2}{c}{\textbf{Algorithm}} & \multicolumn{4}{c}{\textbf{Classification}} & \multicolumn{4}{c}{\textbf{Topic coherence and diversity}} \\
\cmidrule(l{0pt}r{4pt}){2-5}
\cmidrule(l{4pt}r{4pt}){7-8}
\cmidrule(l{4pt}r{4pt}){9-12}
\cmidrule(l{4pt}r{0pt}){13-16}
 & $D$ & \textbf{\#} & $\varnothing$ & \textbf{K} & & $E$ & $\rho$ & \textbf{Acc} & \textbf{Prec} & \textbf{Rec} & \textbf{F1} & \textbf{NPMI} & \textbf{UMass} & $\mathbf{C_V}$ & \textbf{Div} \\
\midrule
\multirow{2}{*}{Banking} 
  & \multirow{2}{*}{10\,003} & \multirow{2}{*}{20} & \multirow{2}{*}{10.76} & \multirow{2}{*}{77} & SPF & 750& 0.001& 0.72 & 0.73 & 0.71 & 0.71 & $\phantom{-}$0.03 & $-$1.97 & 0.44 & 0.12 \\
  &                         &                      &                        &                     & PF  & 750& 0.001& 0.04& 0.04& 0.04& 0.04& $\phantom{-}$0.12& $-$1.70& 0.36& 0.01\\
\cmidrule(lr){1-16}
\multirow{2}{*}{20NG} 
  & \multirow{2}{*}{18\,267}& \multirow{2}{*}{5} & \multirow{2}{*}{91.48}& \multirow{2}{*}{20} & SPF & 550& 0.01& 0.60& 0.62& 0.58& 0.56& $\phantom{-}$0.02& $-$1.93& 0.70 & 0.79\\
  &                         &                    &                        &                     & PF  & 550& 0.01& 0.35& 0.36& 0.34& 0.33& $-$0.01& $-$2.02& 0.65& 0.76\\
\cmidrule(lr){1-16}
\multirow{2}{*}{Ledgar} 
  & \multirow{2}{*}{60\,000} & \multirow{2}{*}{25} & \multirow{2}{*}{54.43} & \multirow{2}{*}{100} & SPF & 550& 0.0005& 0.61 & 0.52 & 0.58 & 0.51 & $\phantom{-}$0.05 & $-$2.04 & 0.71 & 0.48 \\
  &                          &                     &                        &                      & PF  & 550& 0.0005& 0.02& 0.02& 0.02& 0.02& $\phantom{-}$0.13& $-$1.39& 0.51& 0.02\\
\cmidrule(lr){1-16}
\multirow{2}{*}{DBPedia} 
  & \multirow{2}{*}{559\,975}& \multirow{2}{*}{25} & \multirow{2}{*}{24.74} & \multirow{2}{*}{14} & SPF & 250& 0.01& 0.84 & 0.85 & 0.84 & 0.84 & $\phantom{-}$0.19 & $-$2.45 & 0.78 & 0.95 \\
  &                           &                     &                        &                      & PF  & 250& 0.01& 0.39& 0.39& 0.39& 0.38& $\phantom{-}$0.13& $-$2.70& 0.63& 0.76\\
\bottomrule
\end{tabular}
\end{adjustbox}
\caption{Evaluation of model performance across different datasets. For each dataset, we report key corpus characteristics (\# indicates the number of seed terms per topic, $\varnothing$ denotes the average number of words per document), algorithm settings, classification performance (Accuracy -- Acc, Precision -- Prec, Recall -- Rec, F1-score -- F1), topic coherence (NPMI, UMass, $C_V$) and diversity (Div) for both SPF and PF topic models.}
\label{tab:exp_data}
\end{table}

We set the number of topics to the number of categories provided for each dataset (i.e., 77 for Banking, 14 for DBPedia, 20 for 20NG, and 100 for Ledgar). We initialize the hyperparameters of the gamma priors as outlined in Section~\ref{sec:SPF} and train both models with learning rates and number of epochs, specifically selected for each corpus, closely monitoring the convergence of the ELBO. The settings used and results obtained are reported in Table~\ref{tab:exp_data}. The experiments are run on the hardware described in Section~\ref{sec:computational_details}. For coherence and topic diversity evaluation, we use the top-10 ranked words per topic.

SPF achieves strong predictive performance across all datasets, with the accuracy scores varying between 0.60 and 0.84. A particularly high accuracy score is obtained for the large-scale DBPedia dataset ($0.84$), which has 14 categories. Also the accuracy score of $0.72$ obtained for 
the Banking dataset is impressive, in particular given the 77 categories to which one assigns. This excellent classification performance indicates that SPF maintains competitive results, demonstrating robustness to variations in corpus size, document length, number of topics, and number of seed terms. In contrast, the PF model, which is completely unsupervised, results consistently in lower classification performance, with accuracy scores ranging from $0.02$ on Ledgar to $0.39$ on DBPedia. This highlights the substantial benefit of incorporating minimal domain knowledge through seed terms, as in SPF. 

The topic coherence and diversity scores reveal an expected trade-off. For example, DBPedia, which yields the best classification performance for SPF, also achieves the highest topic diversity ($0.95$) and $C_V$ coherence ($0.78$), while scoring poorly on UMass ($-2.45$). PF, on the other hand, tends to produce slightly higher NPMI and UMass scores in some settings (e.g., Banking), likely due to its unsupervised optimization of topic structure rather than predictive alignment. However, this comes at the cost of classification performance. Overall, while PF occasionally produces more coherent or diverse topic structures according to select metrics, SPF clearly outperforms it in classification tasks. The results confirm that even limited supervision, as provided by seed terms, significantly enhances predictive accuracy with only a modest impact on topic quality.

\color{black}

\section{Discussion}\label{sec:conclusion}

Traditional topic models often struggle to align the latent topics they derive with pre-specified concepts of interest \citep[see, e.g.,][]{eshima2023keyword}. To address these limitations, we extend the PF topic model with a seeded approach. The seeded approach guides the inference of topics, avoiding the need for manual labeling, but also enables the use of topic models for text classification where labeled text data are not available but the classes for categorization are readily characterized by a set of relevant words. Seeding modifies the prior distribution of the topic-term distributions by assigning higher rates a-priori to the relevant words associated with their respective topics.  

Our empirical findings demonstrate that integrating domain knowledge into the model specification significantly enhances the capability of topic models to extract meaningful topic-term intensities, thereby improving the understanding of topics. Additionally, by applying a Naive Bayes classifier based on the fitted document-topic distributions, we are able to classify documents automatically. Experiments on datasets with known categorizations reveal that the SPF topic model achieves superior classification performance compared to alternative seeded probabilistic topic modeling approaches. By combining the computational efficiency of VI techniques with the prior knowledge of domain experts in a PF framework, SPF enables a robust and effective system for document classification. This synergy improves the overall quality and utility of the classification process, making it more reliable and actionable for a wide range of applications.  

SPF relies on the bag-of-words assumption to allow straightforward inclusion of domain knowledge and efficient estimation. SPF, however, might benefit, in particular, from recent advances in deep learning methods, including transformer or Mamba models, by exploiting their capabilities for the improved derivation of sets of terms to be used as seed words, see \citet{meng2020discriminative} and \citet{zhang2023effective}.

\if0\blind
{
\section*{Acknowledgements}
The authors gratefully acknowledge support from the Jubil\"aumsfonds of the Oesterreichische Nationalbank (OeNB, grant no.~18718).
}
\fi
\if1\blind
{

}
\fi

\printbibliography

\end{document}